\documentclass[runningheads]{llncs}
\usepackage{graphicx}
\usepackage{amsmath}
\usepackage{amssymb}
\usepackage{booktabs}

\usepackage[normalem]{ulem}
\usepackage{setspace}
\usepackage[vlined, ruled, boxed, linesnumbered]{algorithm2e}
\SetKw{Continue}{continue}
\SetKw{Break}{break}
\SetKwFor{ForAll}{for all}{do}{end}
\SetKwFor{ParForAll}{for all}{in parallel with index $j$ do}{end}
\SetKwFor{ParForAllNoIndex}{for all}{in parallel do}{end}
\SetKwRepeat{Do}{do}{while}
\SetKwInOut{Parameter}{Parameters}
\SetKwComment{Comment}{$\triangleright$\ }{}

\usepackage{mathtools}
\DeclarePairedDelimiter{\ceil}{\lceil}{\rceil}

\usepackage{wrapfig}

\newcommand{\R}{\ensuremath{\mathbb{R}}}

\newcommand{\compactAlgorithm}{\setstretch{0.1}}

\newcommand{\PIRK}{{\tt PIRK}}
\newcommand{\TIRA}{{\tt TIRA}}
\newcommand{\pFaces}{{\tt pFaces}}

\usepackage[style=numeric-comp, sorting=none]{biblatex}
\begin{document}
\title{\texttt{PIRK}: Scalable Interval Reachability Analysis for High-Dimensional Nonlinear 
Systems}
\titlerunning{PIRK: Parallel Reachability}
\author{Alex Devonport$^{*,}$\inst{1} \and
Mahmoud Khaled$^{*,}$\inst{2} \and
Murat Arcak\inst{1} \and 
Majid Zamani\inst{3,4}}
\authorrunning{Devonport et al.}
\institute{
    University of California, Berkeley, Berkeley CA, USA\\
    \email{\{alex\_devonport, arcak\}@berkeley.edu}
    \and
    Technical University of Munich, Germany \\
    \email{khaled.mahmoud@tum.de}
    \and 
    University of Colorado, Boulder, Boulder CO, USA\\
    \email{majid.zamani@colorado.edu}
    \and
    Ludwig Maximilian University, Munich, Germany\\
    $^*$Both authors have contributed equally.
}
\maketitle              %
\begin{abstract}
Reachability analysis is a critical tool for the formal verification of dynamical systems and the synthesis of controllers for them.
Due to their computational complexity, many reachability analysis methods are restricted to systems with relatively small dimensions. 
One significant reason for such limitation is that those approaches, and their implementations, are not designed to leverage parallelism.
They use algorithms that are designed to run serially within one compute unit and they can not utilize widely-available high-performance 
computing (HPC) platforms such as many-core CPUs, GPUs and Cloud-computing services.

This paper presents \texttt{PIRK}, a tool to efficiently compute reachable sets 
for general nonlinear systems of extremely high dimensions.
\texttt{PIRK} can utilize HPC platforms for computing reachable sets for 
general high-dimensional non-linear systems.
\texttt{PIRK} has been tested on several systems, with state 
dimensions ranging from ten up to 4 billion.
The scalability of \texttt{PIRK}'s parallel implementations is found to be 
highly favorable.

\end{abstract}

\section{Introduction}

Applications of safety-critical cyber-physical systems (CPS) are growing due to emerging IoT technologies and the increasing availability of efficient computing devices.
These include smart buildings, traffic networks, autonomous vehicles, truck platooning, and drone swarms, which require reliable bug-free software that perform in real-time and fulfill design requirements.
Traditional simulation/testing-based strategies may only find a small percentage of the software defects and the repairs become much costly as the system complexity grows.
Hence, in-development verification strategies are favorable since they reveal the faults in earlier stages, and guarantee that the designs satisfy the specifications as they evolve through the development cycle.
Formal methods offer an attractive alternative to testing- and simulation-based approaches, as
they can verify whether the specifications for a CPS are satisfied for all possible behaviors from a set of the initial states of the system.
\emph{Reachable sets} characterize the states a system can reach in a given time 
range, starting from a certain initial set and subjected to certain inputs.
Reachable sets play an important role in several formal methods-based approaches to the verification and controller synthesis. An important example of this is \emph{abstraction-based} synthesis
\cite{majid, belta, tabuada, gunther2},
in which reachable sets are used to construct a finite-state 
``abstraction'' which is then used for formal synthesis.

Computing the exact reachable set is generally not possible. For example, even 
for the case of discrete-time LTI systems it is not known whether the
exact reachable set is computable in many important applications 
\cite{fijalkow2019decidability}.
Therefore, most practical methods resort to computing over-approximations or under-approximations of the reachable set, depending on the desired guarantee. 
Computing these approximations to a high degree of accuracy is still a computationally intensive task, particularly for high-dimensional systems. 
Many software tools have been created to address the various challenges of 
approximating reachable sets. Each of these tools uses different methods and 
leverages different system assumptions to achieve different goals related to 
computing reachable sets. For example, \texttt{CORA} 
\cite{althoff2015introduction}
and \texttt{SpaceEx} \cite{FrehseLGDCRLRGDM11}
tools are designed to compute reachable sets of high accuracy for very general classes of nonlinear systems, including hybrid ones.
Some reachability analysis methods rely on specific features of dynamical 
systems, such as linearity of the dynamics or sparsity in the interconnection 
structure \cite{bak2019numerical}.
This allows computing the reachable sets in shorter time or for relatively high-dimensional systems.
However, it limits the approach to smaller classes of applications, less practical specifications, or requires the use of less accurate (e.g., linearized) models.
Some examples of toolboxes that make these kinds of assumptions are the \texttt{Ellipsoidal Toolbox} \cite{kurzhanskiy2006ellipsoidal},
which assumes that system is linear (though it is allowed to be time-varying and subject to disturbance), and \texttt{SAPO} \cite{dreossi2017sapo}, which assumes polynomial dynamics.
Other methods attack the computational complexity problem by computing reachable set approximations from a limited class of set representations.

An example of limiting the set of allowed overapproximations are \emph{interval reachability} methods, in which reachable sets are approximated by Cartesian products of intervals. 
Interval reachability methods allow for computing the reachable sets of very general non-linear and high-dimensional systems in a short amount of time. 
They also pose mild constraints on the systems under consideration, usually only requiring some kind of boundedness constraint instead of a specific form for the system dynamics. Many reachability tools that are designed to scale well with state dimension focus on interval reachability methods: these include
\texttt{Flow$^*$} \cite{chen2013flow},
\texttt{CAPD} \cite{capdwebsite},
\texttt{C2E2} \cite{duggirala2015c2e2},
\texttt{VNODE-LP} \cite{nedialkov2011implementing},
\texttt{DynIbex} \cite{sandretto2016validated},
and \TIRA{} \cite{meyer2019tira}.

Another avenue by which reachable set computation time can be reduced, which we believe has not been sufficiently explored, is the use of parallel computing.
Although most reachability methods are presented as serial algorithms, many of them have some inherent parallelism that can be exploited.
One example of a tool that exploits parallelism is \texttt{XSpeed} \cite{gurung2015xspeed},
which implements a parallelized version of a support function-based reachability method. However, this parallel method is limited to linear systems, and in some cases only linear systems with invertible dynamics. Further, the parallelization is not suitable for massively parallel hardware: only some of the work (sampling of the support functions) is offloaded to the parallel device, so only a relatively small number of parallel processing elements may be employed.

In this paper, we investigate the parallelism for three interval reachability analysis methods and introduce \PIRK, the Parallel Interval Reachability Kernel. PIRK uses \emph{simulation-based} reachability methods
\cite{huang2012computing, julius2009trajectory, maidens2014reachability}, 
which compute rigorous approximations to reachable sets by integrating one or more systems of ODEs. 
The three simulation-based methods implemented in \PIRK{} are
(1) the contraction/growth bound method, 
(2) the mixed monotonicity method, and 
(3) a Monte Carlo based method, 
allowing for different tradeoffs between conservatism and computation speed, as 
well as different forms of problem data.

\PIRK{} is developed in {\tt C++} and {\tt OpenCL} as an open-source kernel for 
\pFaces{} \cite{khaled2019pfaces}, 
a recently introduced acceleration ecosystem.
Having a reachability tool written in \texttt{OpenCL} using \texttt{pFaces} has 
several benefits. 
First, it can be run on any massively parallel computation platform: many-core CPUs, GPUs, and hardware accelerators from any vendor, as well as cloud-based services like AWS.
Second, as a kernel on top of pFaces, PIRK benefits from its representational state transfer (REST) API.
The REST API allows PIRK to be deployed remotely, on any HPC platform, as a web application, serving reachability analysis as a service.
Other tools can connect to \PIRK's RESTful interface and request solutions to reachability analysis problems.
\PIRK{} then utilizes the HPC platform to efficiently solve the problem and responds with the solution.

The analyst looking to use a reachability analysis tool for formal verification may choose from an abundance of options, as our brief review has shown. What PIRK offers in this choice is a tool that allows for massively parallel reachability analysis of high-dimensional systems with an API to easily interface with other tools.
To the best of our knowledge, \PIRK{} is the first and the only tool that can compute reachable sets of general non-linear systems with dimensions beyond the billion.
As we show later in Section \ref{sec:examples}, \PIRK{} computes the reachable set for a traffic network example with 4 billion dimension in only 44.7 minutes using a 96-core CPU in Amazon AWS Cloud.
\section{Interval Reachability Analysis}
Consider a nonlinear system with dynamics 
$\dot{x} = f(t,x,p)$
with state $x\in\R^{n}$,
a set of initial states $\mathcal{X}_0$, 
a time interval $[t_0,t_1]$,
and a set of time-varying inputs $\mathcal{P}$ defined over $[t_0,t_1]$.
Let $\Phi(t;t_0,x_0,p)$ denote the state of the system, at time $t$, of the 
trajectory beginning at time $t_0$ at initial state $x_0$ under input $p$.
We assume the systems are continuous-time. 
While \PIRK{} cannot 
be used on systems with general discrete transitions, it can simulate switched 
systems by using a system input as a switching signal that encodes transition 
times and state.

\begin{figure}[t]
    \centering
    \includegraphics[width=0.5\textwidth]{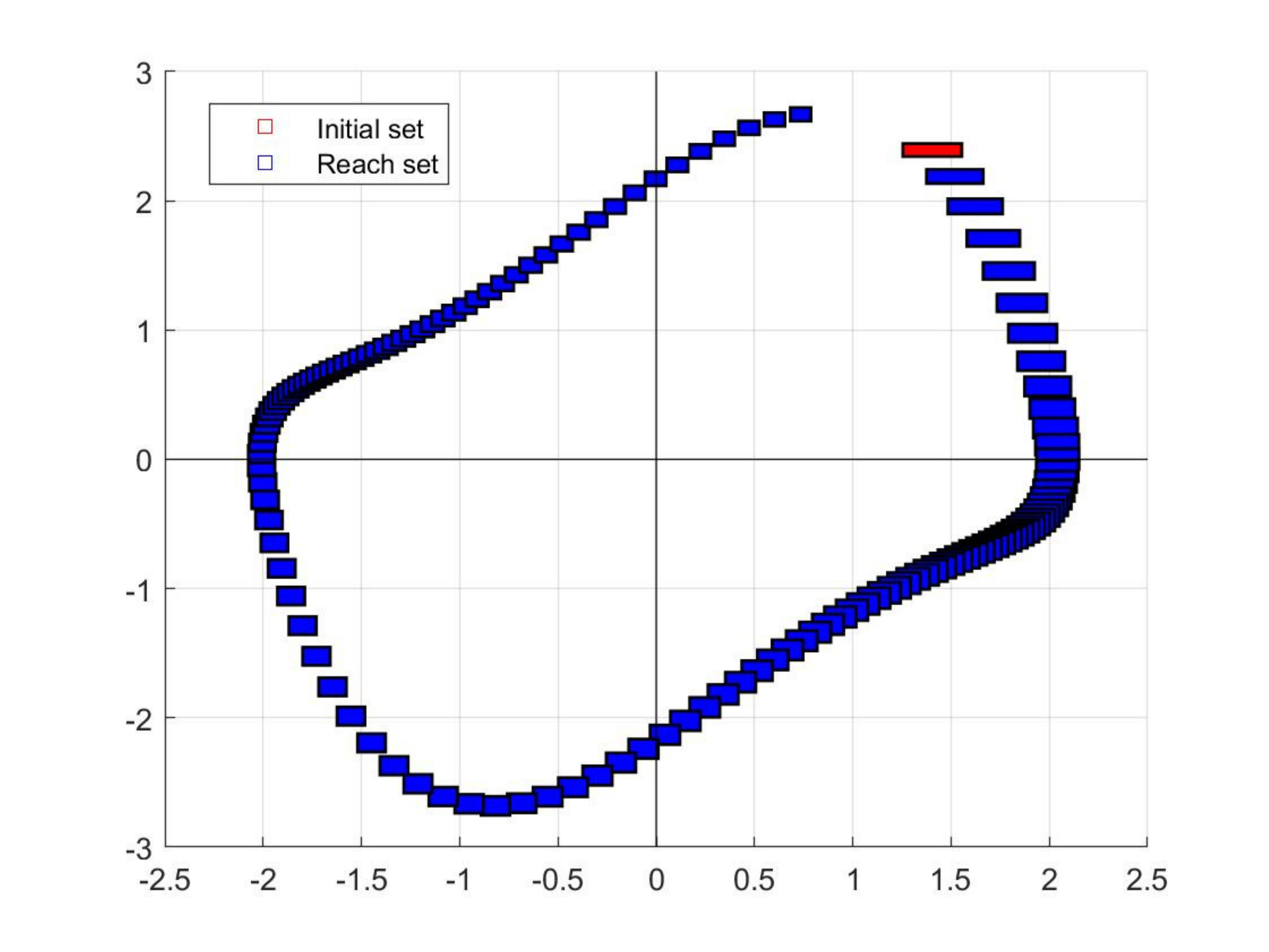}
    \caption{An example of an Interval Reachability problem for the Van der Pol oscillator, a nonlinear system with two states. The red rectangle is the interval initial set. The blue rectangles are interval reachable sets for several final times $t_1$.}
    \label{fig:example reachable set}
\end{figure}

The finite-time forward reachable set is defined as
\begin{equation}
\nonumber
 R_{[t_0,t_1]} = \{\Phi(t_1;t_0,x,p) | x\in\mathcal{X}_0,p\in\mathcal{P}\}.
\end{equation}
Computing an exact reachable set is typically an impossible or at best intractable task, so we instead aim to compute an approximation $\hat{R}_{[t_0,t_1]}$ to the reachable set, either an over-approximation (so that $R_{[t_0,t_1]} \subset \hat{R}_{[t_0,t_1]}$) or an under-approximation (so that $\hat{R}_{[t_0,t_1]}\subset R_{[t_0,t_1]}$).
An example of several finite-time reachable sets is shown in Figure \ref{fig:example reachable set}.

For the problem of \emph{interval} reachability analysis, there are a few more constraints on the problem structure. 
An \emph{interval set} is a set of the form $[\underline{a},\overline{a}]=\{a: \underline{a} \le a \le \overline{a} \}$, where $\le$ denotes the usual partial order on real vectors, that is the partial order with respect to the positive orthant cone. The vectors $\underline{a}$ and $\overline{a}$ are, respectively, the lower and upper bounds of the interval set.
An interval set can alternatively be described by its center $a^*=\frac{1}{2}(\overline{a}+\underline{a})$
and half-width
$[a]=\frac{1}{2}(\overline{a}-\underline{a})$.
In interval reachability analysis,
the initial set must be an interval, and inputs must be constants drawn from an interval set, i.e. $p(t)=p$, where $p\in[\underline{p},\overline{p}]$, and the reachable set approximation must also be an interval.
Furthermore, certain methods for computing interval reachable sets require further restrictions on the system dynamics, such as the state and input Jacobian matrices being bounded or sign-stable.

\subsection{Methods to compute interval reachable sets}
\PIRK{} computes interval reachable sets using three different methods, allowing for different levels of tightness and speed, and which allow for different amounts of additional problem data to be used. 

The \emph{Contraction/Growth Bound} method \cite{gunther2, kapela2007lohner, fan2016locally} computes the reachable set using component-wise contraction properties of the system. 
This method may be applied to input-affine systems of the form
$ \dot{x}=f(t,x) + p$.
The growth and contraction properties of each component of the system are first characterized by a \emph{contraction matrix} $C$. The contraction matrix is a 
component-wise generalization of the matrix measure of the Jacobian \cite{maidens2014reachability, arcak2018simulation},
satisfying the following:
$ C_{ii} \ge J_{x,ii}(t,x,p^*) $ for diagonal elements, and
$ C_{ij} \ge |J_{x,ij}(t,x,p^*)|$ for off-diagonal elements. 
The method constructs a reachable set over-approximation by separately 
establishing its \emph{center} and \emph{half-width}. The center is found by 
simulating the trajectory of the center of the initial set, that is as 
$\Phi(t_1;t_0,x^*,p^*)$. The half width is found by integrating the \emph{growth 
dynamics} 
$ \dot{r}= g(r,p) = Cr + [p], $
where $r$ denotes the half-width, over the time range $[t_0,t_1]$ with initial condition $r(t_0)=[x]$.

The \emph{Mixed-Monotonicity} method \cite{coogan2017formal} computes the reachable set by separating the increasing and decreasing portions of the system dynamics in an auxiliary system called the \emph{embedding system} whose state dimension is twice that of the original system
\cite{gouze1994monotone}.
The embedding system is constructed using a \emph{decomposition function}
$d(t,x,p,\hat{x},\hat{p})$, which encodes the increasing and decreasing parts of the system dynamics and satisfies $d(t,x,p,x,p)=f(t,x,p)$. The embedding system dynamics are then defined as
\begin{equation}
\nonumber
\label{eq:embedding system}
\begin{bmatrix} \dot{x} \\ \dot{\hat{x}} \end{bmatrix}
= h(x,p,\hat{x},\hat{p})=
\begin{bmatrix} d(t,x,p,\hat{x},\hat{p})\\ d(t,\hat{x},\hat{p}, x,p) \end{bmatrix}.
\end{equation}
For states having $x=\hat{x}$, (\ref{eq:embedding system}) becomes two copies of the original system dynamics by the properties of the decomposition function: this is the sense in which the embedding system ``embeds'' the original system dynamics.

The evaluation of a single trajectory of the embedding system can be used to find a reachable set over-approximation for the original system.
Specifically, consider the solution of (\ref{eq:embedding system}) over the interval $[t_0,t_1]$ with the initial condition
$[ x_0~\hat{x}_0]^T = [\underline{x}~\overline{x}]^T,$
and subject to the constant inputs $p=\underline{p}$, $\hat{p}=\overline{p}$.
Then, the integrated final state $[x_1~\hat{x}_1]^T$ serves as an over-approximation of the reachable set of the original system dynamics, that is $R_{[t_0,t_1]} \subset [x_1,\hat{x}_1]$.   

The \emph{Monte Carlo} method computes a probabilistic approximation to the reachable set by evaluating the trajectories of a finite number of \emph{sample points} in the initial set and input set, and selecting the smallest interval that contains the final points of the trajectories.
More formally, we take a probability distribution whose support is the initial set (for example, the uniform distribution over the initial set), and draw $m$ independent and identically distributed (iid) sampled initial states $x_0^{(i)}$.
Each initial state is integrated over $[t_0,t_1]$ to yield a final state 
$x_1^{(i)}$. The interval reachable set is then approximated by the 
elementwise minimum and maximum of the $x_1^{(i)}$.
This approximation satisfies a probabilistic guarantee of correctness, provided 
that enough sample states are chosen \cite{devonport2019datadriven}. Let 
$[\underline{R},\overline{R}]$ be the approximated reachable set,
$\epsilon, \delta \in (0,1)$, and
$m \ge (\frac{2n}{\epsilon})\log\left(\frac{2n}{\delta} \right).$
Then with probability $1-\delta$, the approximation $[\underline{R},\overline{R}]$ satisfies
$P(R_{[t_0,t_1]} \backslash [\underline{R},\overline{R}]) \le \epsilon,$
where $P(A)$ denotes the probability that a sampled initial state will yield a final state in the set $A$, and $\backslash$ denotes set difference.

\section{Parallelization}
The bulk of the computational work in each method is spent in ODE integration.
Hence, the most effective approach by which to parallelize the three methods is to design a parallel ODE integration method.

There are several available methods for parallelizing the task of ODE integration. 
Several popular methods for parallel ODE integration are parallel extensions of Runge-Kutta integration methods, which are the most popular serial methods for ODE integration. 
For these methods, the parallelization may be done in essentially two different ways \cite{solodushkin2016parallel}.
First, they may be parallelized \emph{across space}, in which case the the computations associated to each state variable will be done in parallel. This allows for as many parallel computation elements as there are state variables.
Second, they may be parallelized \emph{across time}, in which case the time interval of integration is split up into several parts, and the solution on each sub-interval is computed in parallel. 
The computations for each sub-interval may be solved iteratively using a two-point shooting method, as in the Parareal algorithm \cite{gander2007analysis}.
In contrast to parallelization across space, parallelization across time can be scaled to an arbitrary degree, as any number of sub-intervals may be used. However, the iterative solution by shooting introduces additional computations.
To avoid this additional overhead, \PIRK{} uses parallelization across space. Since \PIRK{}'s goal is to compute reachable sets for extremely high-dimensional systems, parallelization across space allows for a sufficient degree of parallelization in most cases.

\subsection{Parallelization of the Runge-Kutta scheme}
\begin{algorithm}[t]
    \compactAlgorithm
    \SetAlgoLined
    \caption{State-parallelized fourth-order Runge-Kutta scheme for ODE integration.}
    \label{alg:RKparallel}
    \KwIn{Initial state $x_0$; input $p$; and initial and final times $t_0$,$t_1$.}
    \Parameter{State dimension $n$; time step size $h$ and ODE right-hand side function $d(t,x,p)$.}
    \KwOut{Final state $x_1$ as the solution to the ODE at time $t_1$.}
    
    \ParForAllNoIndex{$i \in \{1,\dotsc,n\}$} {
        $k_{0,i}=k_{1,i}=k_{2,i}=k_{3,i}=\text{tmp}_i=0$; 
        $x_{1,i} = x_{0,i}$\;
    }    
    \For{$t \in \{t_0, t_0+h,\dotsc,t_1-h,t_1\}$} {
        \ParForAllNoIndex{$i \in \{1,\dotsc,n\}$} {
            $k_{0,i}=d_i(t, x_1, p)$;
            $\text{tmp}_i=x_1+\frac{h}{2}k_{0,i}$\;
        }
        \ParForAllNoIndex{$i \in \{1,\dotsc,n\}$} {
            $k_{1,i}=d_i(t+\frac{h}{2}, \text{tmp}, p)$;
            $\text{tmp}_i=x_1+\frac{h}{2}k_{1,i}$\;
        }
        \ParForAllNoIndex{$i \in \{1,\dotsc,n\}$} {
            $k_{2,i}=d_i(t+\frac{h}{2}, \text{tmp}, p)$;
            $\text{tmp}_i=x_1+\frac{h}{2}k_{2,i}$\;
        }
        \ParForAllNoIndex{$i \in \{1,\dotsc,n\}$} {
            $k_{3,i}=d_i(t+h, \text{tmp}, p)$;
            $\text{tmp}_i=x_1+\frac{h}{2}k_{2,i}$\;
            $x_{1,i}=x_{1,i}+\frac{1}{6}\left( k_{0,i}+2k_{1,i}+2k_{2,i}+k_{3,i} \right)$\;
        }
    }
\end{algorithm}	
Algorithm \ref{alg:RKparallel} shows a parallelization across space of the $n$-dimensional fourth-order Runge-Kutta scheme for ODE integration.
It starts with a parallel initialization (step 1) of the internal variables ($k_{0}$, $k_{1}$, $k_{2}$, $k_{3}$, and $\text{tmp}$) and the final state $x_1$.
Then, for each quantized time step $t$, as demonstrated by the \texttt{for} loop in step 4, four parallel sections (starting with steps 5, 8, 11 and 14) are executed on all available processing elements (PEs).
Each parallel section is responsible for computing intermediate component-wise values of the integration and storing them in the memory spaces controlled by the corresponding internal variables.
By the end of the last parallel section, step 17, the component-wise values of the solution to the ODE are computed for the current time step $t$.

In parallelization across space, the separation of integration results in less synchronization overhead between the threads executing the parallel code.
All that is required is to make sure that the threads are synchronized after each of the parallel sections, which is already achieved in the algorithm using separate parallel \texttt{for} loops.
This data-parallel approach fits massively-parallel compute systems, such as GPUs and clusters in the Cloud.

For a system with $n$ dimensions and assuming all PEs have uniform access time to the memory space, Algorithm \ref{alg:RKparallel} scales linearly as the number of PEs (denoted by $P$) increases.
In a computer with a single PE (i.e., $P = 1$), the algorithms reduces to the original serial algorithm.
Let $T$ be the time needed to run the algorithm on such single-PE computer.
Then, if a parallel computer has $P$ PEs of same type, where $P \le n$, users may expect a speedup in time of up to ideally $P$ times.
Here, each PE will be responsible for computing $n/P$ components of the state vector. 
Using $P$ number of PEs, where $P > n$, is a waste of computation resources since one or more PEs will be left idle.
For fixed initial and final times, $t_0$ and $t_1$, the time complexity of the algorithm is $O(\frac{n}{P})$.

\subsection{Parallelization of the Methods}

\begin{algorithm}[t]
    \compactAlgorithm
    \SetAlgoLined
    \caption{Parallel contraction/growth bound method.}
    \label{alg:GBParallel}
    \KwIn{System $\Sigma$, interval initial set $\mathcal{X}_0$ ; interval input set $\mathcal{P}$; and initial and final times $t_0$,$t_1$.}
    \KwOut{Reachable set $\hat{R}_{[t_0,t_1]}$.}
    \ParForAllNoIndex{$i \in \{1,\dotsc,n\}$} {
        Set $c_{0,i}$ as the center of the interval $\mathcal{X}_{0,i}$\;
        Set $r_{0,i}$ as the radius of the interval $\mathcal{X}_{0,i}$\;
    }
    Run Algorithm \ref{alg:RKparallel} with $c_0$ as initial state, $f$ as right-hand side function and the center of $\mathcal{P}$
    as input\;
    Store the result of Algorithm \ref{alg:RKparallel} in $c_1$\;
    Run Algorithm \ref{alg:RKparallel} with $r_0$ as initial state, $g$ as right-hand side function, and
    the half-width of $\mathcal{P}$
    as input\;
    Store the result of Algorithm \ref{alg:RKparallel} in $r_1$\;
    \ParForAllNoIndex{$i \in \{1,\dotsc,n\}$} {
        Set $\hat{R}_{[t_0,t_1],i} = [c_{1,i}-r_{1,i},\;c_{1,i}+r_{1,i}]$ \;
    }    
\end{algorithm}	

Algorithm \ref{alg:GBParallel} presents the parallel version of the contraction/growth bound method.
Algorithm \ref{alg:GBParallel} uses
Algorithm \ref{alg:RKparallel} twice.
First, it is used to compute the solution of the system's 
ODE $f$ for the center of the initial set $I_0$.
Then, it is used to compute the growth/contraction of the initial set 
$I_0$ by solving the ODE $g$ of the growth dynamics.
Since this method uses Algorithm \ref{alg:RKparallel} directly, its time 
complexity is also $O(\frac{n}{P})$, for fixed $t_0$ and $t_1$.

\begin{algorithm}[t]
    \compactAlgorithm
    \SetAlgoLined
    \caption{Parallel mixed monotonicity method.}
    \label{alg:MMParallel}
    \KwIn{System $\Sigma$, interval initial set $\mathcal{X}_0$; decomposition function $g$; interval input set $\mathcal{P}$; and initial and final times $t_0$,$t_1$.}
    \KwOut{Reachable set $\hat{R}_{[t_0,t_1]}$.}
    
    \ParForAllNoIndex{$i \in \{1,\dotsc,n\}$} {
        Set $x_{0,i}$ as the lower bound of the interval $\mathcal{X}_{0,i}$\;
        Set $\hat{x}_{0,i}$ as the upper bound of the interval $\mathcal{X}_{0,i}$\;
        Set $p_i$ as the lower bound of the interval $\mathcal{P}_i$\;
        Set $\hat{p}_i$ as the upper bound of the interval $\mathcal{P}_i$\;
    }
    Run Algorithm \ref{alg:RKparallel} with 
    $\begin{bmatrix} x_0 \\ \hat{x}_0 \end{bmatrix}$ 
    as initial state and 
    $h$ as defined in (\ref{eq:embedding system}) as right-hand side function and $\begin{bmatrix} p \\ \hat{p} \end{bmatrix}$ as input\;
    Store the result of Algorithm \ref{alg:RKparallel} in $x_1$ and $\hat{x}_1$\;

    \ParForAllNoIndex{$i \in \{1,\dotsc,n\}$} {
        Set $\hat{R}_{[t_0,t_1],i} = [x_{1,i},\hat{x}_{1,i}]$ \;
    }    
\end{algorithm}	

Algorithm \ref{alg:MMParallel} presents the parallel version of the method based 
on the mixed monotonicity.
The parallelized implementation of the mixed-monotonicity method uses only one 
call to Algorithm \ref{alg:RKparallel} to integrate the embedding system.
First, the upper and lower bound vectors of the initial sets are extracted in 
parallel 
Then, Algorithm \ref{alg:RKparallel} is called once but with an augmented vector 
constructed by the upper and lower bounds, duplicated ODE right-hand side and 
inputs.
This is similar to running Algorithm \ref{alg:RKparallel} twice, 
which means that the mixed-monotonicity 
method also has a time complexity of $O(\frac{n}{P})$, for fixed $t_0$ and 
$t_1$.
However, since Algorithm \ref{alg:RKparallel} is run on a system of 
dimension $2n$, the mixed-monotonicity method requires twice as much memory as 
the growth bound method.
This limits the algorithm to systems with half the dimensions, compared to Algorithm \ref{alg:GBParallel}, for the same hardware configuration.

\begin{algorithm}[t]
    \compactAlgorithm
    \SetAlgoLined
    \caption{Parallel Monte Carlo simulation method.}
    \label{alg:MCParallel}
    \KwIn{System $\Sigma$, interval initial set $\mathcal{X}_0$ interval input set $\mathcal{P}$; initial and final times $t_0$,$t_1$; probabilistic guarantee parameters $\epsilon$ and $\delta$.}
    \KwOut{Reachable set $\hat{R}_{[t_0,t_1]}$.}
    Set number of samples $m=\ceil*{\frac{2n}{\epsilon}\log\left(\frac{2n}{\delta} \right)}$\;
    \ParForAllNoIndex{$i \in \{1,\dotsc,m\}$} {
        Sample $x_0^{(i)}$ uniformly from $I_0$\;
        Run Algorithm \ref{alg:RKparallel} with $x_0^{(i)}$ as initial state and $f$ as right-hand side function\;
        Store the result of Algorithm \ref{alg:RKparallel} in $x_1^{(k)}$
        \;
    }
    \ParForAllNoIndex{$j \in \{1,\dotsc,n\}$} {
            Set $\hat{R}_{[t_0,t_1],j} = [\underset{i \in \{1 \cdots m\}}{\min} x_{1,j}^{(i)}, \underset{i \in \{1 \cdots m\}}{\max} x_{1,j}^{(i)}] $
    }
\end{algorithm}	

Algorithm \ref{alg:MCParallel} presents the parallelized implementation of the Monte Carlo method. 
Algorithm \ref{alg:MCParallel}uses Algorithm \ref{alg:RKparallel} $m$ times, for $m$ sampled initial states.
The implementation uses two levels of parallelization. 
The first level is a set of parallel threads over the samples used for simulations.
Then, within each thread, another parallel set of threads are launched, as a 
result of calling Algorithm \ref{alg:RKparallel}.
This is realized as one parallel job of $m \times n$ threads.
Consequently, the Monte Carlo method has a complexity of 
$O(\frac{mn}{p})$.
Since only the elementwise minima and maxima of the sampled states need to be 
stored, this method only requires as much memory as the growth bound method.

\section{Working with \PIRK{}}
\PIRK{} is built as a kernel to be launched by the acceleration ecosystem \pFaces{}.
To use \PIRK{}, you should first download and install \pFaces{} (see \pFaces{} guide \cite{pFacesRepo}).
\pFaces{} works with Windows, Linux and MacOS, and \PIRK{} is tested on all of them.
For the sake of demonstration, we assume the readers are using Linux or MacOS.
Having \pFaces{} installed, download and build \PIRK{} by running the following two command lines:
\vspace{0.5em}

{\setlength{\parindent}{1em} \tt\textbf{ \$ git clone TOOL\_URL; cd pFaces-PIRK; make}}

\vspace{0.5em}
where {\tt TOOL\_URL} is the web-address of the tool's repository. 
Now, \PIRK{} is ready and you can launch it with any of the provided examples using the command:
\vspace{0.5em}

{\setlength{\parindent}{1em} \tt\textbf{ \$ pfaces -G -d 1 -k pirk -cfg CONFIG\_FILE}}

\vspace{0.5em}
where {\tt CONFIG\_FILE} is the path to the example's configuration file. 
For instance, to run the traffic network example, which is discussed later in Section \ref{sec:examples}, 
you may replace {\tt \textbf{CONFIG\_FILE}} with {\tt \textbf{./examples/ex\_n\_links/ex\_n\_link.cfg}}.
This simply asks \pFaces{} to run \PIRK{} (-k pirk) on the first (-d 1) GPU (-G) device and use the configuration file 
{\tt ex\_n\_link.cfg} as input.
To add an example, users will need to create a configuration file describing the system's information, initial and final times, accuracy, and other parameters.
They also need to provide the system dynamics in a separate file as a {\tt C}-language function.
For each of the implemented methods, users may provide different input to \PIRK{}.

\section{Case Studies}
\label{sec:examples}
In each of the case studies to follow, we report the time it takes \PIRK{} to compute reachable sets for systems of varying dimension using all three of its methods on a variety of parallel computing platforms. We perform some of the same tests using the serial tool TIRA, to measure the speedup gained by \PIRK's ability to use massively parallel hardware.

We set a time limit of 1 hour for all of the targeted case studies, and report 
the maximum dimensions that could be reached under this limit.
We use four AWS machines for the computations with \PIRK:
{\tt m4.10xlarge} which has a CPU with 40 cores, {\tt c5.24xlarge} which a CPU with 96 cores, {\tt g3.4xlarge} which has a GPU with 2048 cores, and {\tt p3.2xlarge} which has a GPU with 5120 cores.
For the computations with \TIRA, we used a machine with a 3.6 GHz Intel i7 CPU. 

The two most common sources of extremely high-dimensional systems are 
discretizations of continuum models (e.g. discretized PDEs) and swarms of 
independent agents. Three of the following case studies come from these 
sources. such systems naturally to have sparse dynamics, which \PIRK{} 
uses to speed up ODE integration and reduce memory use. However, sparsity is not 
a requirement for \PIRK{} to be effective, and we have used systems 
with dense dynamics for test cases as well.

\begin{figure*}[t]
\begin{center}
    \hspace*{-2.6cm}
    \includegraphics[width=1.38\textwidth]{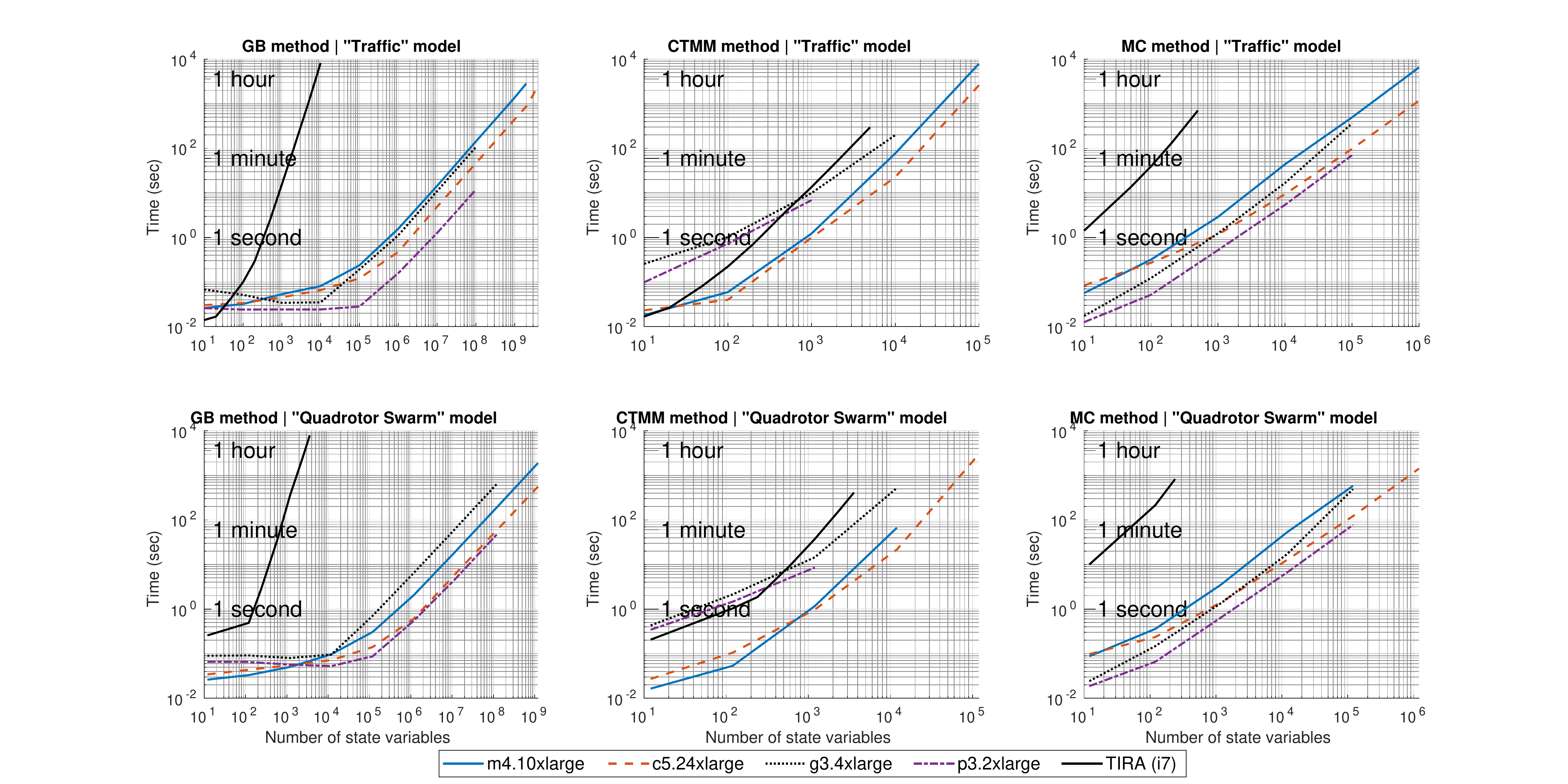}
    \caption{Logarithmic plots of the results for speed tests of the traffic 
model (first row) and the quadrotor swarm (second row). Speed test results for 
the serial interval reachability toolbox TIRA are also shown for the traffic 
model.}
    \label{fig:speed_results}
    \end{center}
\end{figure*}

\subsection{$n$-link Road Traffic Model}

We consider the road traffic analysis problem reported in \cite{coogan2018benchmark},
a proposed benchmark for formal controller synthesis. 
We are interested in the density of cars along a single one-way lane. 
The lane is divided into $n$ segments, and the density of cars in each segment is a state variable. 
The continuous-time dynamics for the segment densities are derived from a spatially discretized version of the Cell Transmission Model \cite{coogan2015compartmental}. 
Spatially discretized Cell Transmission Model systems have been used in conjunction with abstraction-based formal control synthesis \cite{coogan2017formal}.
For the single-road case described here, the dynamics are
\begin{align*}
    \dot{x}_0 &= \frac{1}{T}\left( \beta \min(c, vx_{1}, w(\overline{x}-x_{2})/\beta)\right)\\
    \dot{x}_i &= \frac{1}{T}\left( \beta \min(c, vx_{i-1}, w(\overline{x}-x_{i})/\beta) -  \min(c, vx_{i}, w(\overline{x}-x_{i+1})/\beta)\right)\\
    \dot{x}_{n-1} &= \frac{1}{T}\left( \beta \min(c, vx_{i-1}, w(\overline{x}-x_{i})/\beta) - \min(c, vx_{i})/\beta)\right)
\end{align*}
where $x_i$ represents the density of traffic in the $i$th discretized segment.
Since a system of this form can easily be defined for any natural number $n$, this system is a good candidate for testing the effectiveness of the parallelized reachability methods as a function of state dimension.
This is a nonlinear system with sparse coupling between the state variables, 
in the sense that the dynamics for each state variable $x_i$ depends only on 
itself and its ``neighbors'' $x_{i-1}$ and $x_{i+1}$. This means that 
the system's Jacobian matrix is sparse, which
\PIRK{} can use to reduce the used memory.

The results of the speed test are shown in the first row of Figure 
\ref{fig:speed_results}.
The machines {\tt m4.10xlarge} and {\tt c5.24xlarge} reach up to 2 billion and 4 billion dimensions, respectively, using the growth/contraction method, in 47.3 minutes and 44.7 minutes, respectively.
Due to memory limitations of the GPUs, the machines {\tt g3.4xlarge} and {\tt p3.2xlarge} both reach up to 400 million in 106 seconds and 11 seconds, respectively. 

The relative improvement of \PIRK's computation time over \TIRA's is significantly larger for the growth bound method than for the other two. This difference stems from a difference in how each tool computes the half-width of the reachable set from the radius dynamics. \TIRA solves the radius dynamics by computing the full matrix exponential using MATLAB's \texttt{expm}, whereas \TIRA{} directly integrates the dynamics using parallel Runge-Kutta. This caveat applies to section \ref{sec: quadrotor swarm} as well.

\subsection{Quadrotor Swarm}
\label{sec: quadrotor swarm}
The second test system is a swarm of $K$ identical quadrotors.
The dynamics for an individual quadrotor's position and angle are
\begin{align*}
\ddot{p}_n &= \frac{F}{m}(-\cos(\phi)\sin(\theta)\cos(\psi)-\sin(\phi)\sin(\psi))\\
\ddot{p}_e &= \frac{F}{m}(-\cos(\phi)\sin(\theta)\sin(\psi)+\sin(\phi)\cos(\psi))\\
\ddot{h} &= g-\frac{F}{m}\cos(\phi)\cos(\theta)\\
\ddot{\phi} &= \frac{1}{J_x} \tau_\phi\\
\ddot{\theta} &= \frac{1}{J_y} \tau_\theta\\
\ddot{\psi} &= \frac{1}{J_z} \tau_\psi
\end{align*}
where $p_n$ and $p_e$ denote the y-axis (``north'') and x-axis (``east'') position of the quadrotor, $h$ its height, and $\phi$, $\theta$, and $\psi$ its pitch, roll, and yaw angles respectively.
The system dynamics of each quadrotor model are derived in a similar way to the model used in the ARCH competition \cite{immler2018arch}, with the added simplification of a small angle approximation in the angular dynamics and the neglect of Coriolis force terms. 
A derivation of both models is available in \cite{beard2008quadrotor}.
Similar to the n-link traffic model, this system is convenient for scaling: system consisting of one quadrotor can be expressed with 12 states, so the state dimension of the swarm system is $n=12K$.
The results of the speed test are shown in figure \ref{fig:speed_results} (second row).
The machines {\tt m4.10xlarge} and {\tt c5.24xlarge} reach up to 1.8 billion dimensions and 3.6 billion dimensions, respectively, (using the growth/contraction method) in 48 minutes and 32 minutes, respectively.
The machines {\tt g3.4xlarge} and {\tt p3.2xlarge} both reach up to 120 million dimensions in 10.6 minutes and 46 seconds, respectively. 

\subsection{Quadrotor Swarm With Artificial Potential Field}
\label{sec: quadrotor swarm with artificial potential field}

The third test system is a modification of the quadrotor swarm system which adds interactions between the quadrotors. In addition to the base quadrotor dynamics described in Section \ref{sec: quadrotor swarm}, this model augments each quadrotor with an artificial potential field to guide the quadrotors to the origin while avoiding collisions. 

This controller applies a force to the quadrotors that seeks to minimize a \emph{artificial potential} $U$ that depends on the position of all of the quadrotors. The potential applied to each quadotor is intended to repel it from the other quadrotors, and to attract it towards the origin. Each quadrotor will be subject to three potential fields, which separately consider $p_n$, $p_e$, and $h$. This allows us to ensure that the forces resulting from the potential fields are contained in an interval set.

The potential fields for the $i^{th}$ quadrotor are 
\begin{align*}
 U_{i,p_n} &= F_{r}e^{-min_{j\ne i} |p_n^{(i)}-p_n^{(j)}|} + F_{a} |p_n^{(i)}|\\
 U_{i,p_e} &= F_{r}e^{-min_{j\ne i} |p_e^{(i)}-p_e^{(j)}|} + F_{a} |p_e^{(i)}|\\
 U_{i,h} &= F_{r}e^{-min_{j\ne i} |h^{(i)}-h^{(j)}|} + F_{a} |e^{(i)}|,
\end{align*}
where the $(i)$ and $(j)$ superscripts denote state variables belonging to the $i^{th}$ and $j^{th}$ quadrotors respectively.
The magnitude of this potential field depends on the distance from the origin and the position of the nearest quadrotor in the $p_n$, $p_e$, and $h$ directions, so the control action arising from this potential field will tend to guide a quadrotor away from whichever other quadrotor is nearest to it while also guiding it towards the origin.

The forces applied to the system in each direction are just the negative derivatives of the each direction's potential field.
Therefore, the dynamics for quadrotor $i$ are
\begin{align*}
\ddot{p}_n &= \frac{F}{m}(-\cos(\phi)\sin(\theta)\cos(\psi)-\sin(\phi)\sin(\psi)) + F_{i,p_n}\\
\ddot{p}_e &= \frac{F}{m}(-\cos(\phi)\sin(\theta)\sin(\psi)+\sin(\phi)\cos(\psi)) + F_{i,p_e}\\
\ddot{h} &= g-\frac{F}{m}\cos(\phi)\cos(\theta) + F_{i,h}\\
\ddot{\phi} &= \frac{1}{J_x} \tau_\phi\\
\ddot{\theta} &= \frac{1}{J_y} \tau_\theta\\
\ddot{\psi} &= \frac{1}{J_z} \tau_\psi,
\end{align*}
where $F_{i,p_n}$, $F_{i,p_e}$, and $F_{i,h}$ are the forces induced by the artificial potential fields. These have the form
\newcommand{\js}{{j^{*}}}
\begin{align*}
 F_{i,p_n} &= -\frac{\partial U_{i,p_n}}{\partial p_n^{(i)}} =
 F_r \text{sgn}(p_n^{(i)}-p_n^{(\js)})e^{-|p_n^{(i)}-p_n^{(\js)}|} - F_a\text{sgn}(p_n^{(i)})\\
 F_{i,p_e} &= -\frac{\partial U_{i,p_e}}{\partial p_e^{(i)}} =
 F_r \text{sgn}(p_e^{(i)}-p_e^{(\js)})e^{-|p_e^{(i)}-p_e^{(\js)}|} - F_a\text{sgn}(p_e^{(i)})\\
 F_{i,h} &= -\frac{\partial U_{i,h}}{\partial h^{(i)}} =
 F_r \text{sgn}(h^{(i)}-h^{(\js)})e^{-|h^{(i)}-h^{(\js)}|} - F_a\text{sgn}(h^{(i)}),
\end{align*}
where $\js=\text{argmin}_{j\ne i} |x_i-x_j|$ in the $F_{i,x}$ equation, and is defined analogously for the other two. Each force is bounded by the interval $[-(F_r+F_a),F_r+F_a]$. The second derivatives of the potential functions, which appear in the system Jacobian, are be bounded by the interval $[-F_r,F_r]$ almost everywhere. Due to the interaction of the state variables in the $F_{i,p_n}$, $F_{i,p_e}$, and $F_{i,h}$ terms, this system has a dense Jacobian. In particular, at least 25\% of the Jacobian elements will be nonzero for any number of quadrotors.

\begin{table}
    \centering
    \begin{tabular}{l|l|c|c|c|c|c}
        \toprule
        
        &
        No. of&
        Memory&
        \multicolumn{4}{c}{Time (seconds)}\\
        
        Method & 
        States & 
        (MB)& 
        \texttt{m4.10xlarge} & 
        \texttt{c5.24xlarge} & 
        \texttt{g3.4xlarge} & 
        \texttt{p3.2xlarge} \\
        
        \midrule
        GB &    1200     & 2.8       & $\le$ 1.0 & $\le$ 1.0 & 4.3       & $\le$ 1.0 \\
        GB &    12000    & 275.3     & $\le$ 1.0 & $\le$ 1.0 & 47.1      & $\le$ 1.0 \\
        GB &    120000   & 27,473.1  & 69.6      & 68.3      & N/M       & N/M  \\
        \midrule
        MC &    1200     & 45.7      & 1.0       & $\le$ 1.0 & 2.0       & $\le$ 1.0 \\
        MC &    12000    & 457.5     & 56.8      & 23.7      & 233.1     & 40.6 \\
        MC &    120000   & 4577.6    & $\ge$ 2h  & 3091.8    & N/M       & 5081.0 \\    
        \bottomrule
    
    \end{tabular}
    \label{tab:quad_apf_times}
    \caption{Results for running \PIRK{} to compute the reach set of the quadrotors swarm with artificial potential field. ``N/M'' means that the machine did not have enough memory to compute the reachable set.}
\end{table}

Table \ref{tab:quad_apf_times} shows the times of running \PIRK{} using his system within the four machines 
{\tt m4.10xlarge}, {\tt c5.24xlarge}, {\tt g3.4xlarge} and {\tt p3.2xlarge} in Amazon AWS.
Due to the high density of this example, the required memory grrows very fast.
Here, we focus on the growth bound and the Monte-Carlo methods since they require less memory.
For the Monte-Carlo method, we fix the number of samples to 1000 samples.

\PIRK{} manged to compute the reach sets of systems up to 120,000 state variables (i.e., 10,000 quadrotors).
Up to 1,200 states, all machines solve the problems in less than one second.
Some of the machines lack the required memory to solve the problems with large memory requirement 
(e.g., a 27.7 GB memory is required to compute the reach set of the system with 120,000 state variables using the growth bound method).

\subsection{Heat Diffusion}
\label{sec: heat diffusion}
The fourth test system is a model for the diffusion of heat in a 3-dimensional cube. The model is based on on a benchmark used in \cite{bak2019numerical} to test a method for numerical verification of affine systems. 
The solid is a cube of unit length, which is insulated on all sides but one: the non-insulated side can exchange heat with the outside environment, which is assumed to be at zero temperature. 
A portion of the cube is heated at an initial time, and the heat diffuses through the rest of the cube as time progresses, until all of the heat has left through the non-insulated side.
The dynamics for the transfer of heat through the cube are governed by the \emph{heat equation},
a classical linear partial differential equation.
a second-order linear differential equation of the form
\begin{equation*}
    \frac{\partial u}{\partial t} = \alpha \left( 
    \frac{\partial^2 u}{\partial x^2}+\frac{\partial^2 u}{\partial y^2}+\frac{\partial^2 u}{\partial z^2} 
    \right),
\end{equation*}
where $u(t;x,y,z)$ denotes the temperature of the cube at coordinates $x$, $y$, and $z$ at time $t$, and $\alpha$ is the diffusivity of the material of the cube.
A model of the form $\dot{x} = f(t,x,p)$ which approximates the heat 
transfer through the cube according to the heat equation can be obtained by discretizing the cube into an
$\ell\times\ell\times \ell$ grid, yielding a system with $\ell^3$ states.
The temperature at each grid point is taken as a state variable. Each spatial 
partial derivative is replaced with its first-order discrete approximation based 
on the discretized state variables, accounting for boundary conditions. 
Since the heat equation is a linear PDE, the discretized system is linear.

We take a fixed state dimension of $n=10^9$ by fixing $\ell=1000$. 
This example uses a finer time-sampling constant compared to other examples. Integration takes place over $[t_0,t_1]=[0,20]$ with time step size $h=0.02$, leading to 1000 integration steps.
\PIRK{} solves the problem on {\tt m4.10xlarge} in 472 minutes, and in 350.2 minutes on {\tt c5.24xlarge}.

\subsection{Overtaking Maneuver with a Single-Track Vehicle}

This and the remaining case studies focus on models of practical importance which have relatively low state dimension. Although \PIRK{} is designed to perform well on high-dimensional systems, it is also effective at quickly computing reachable sets for lower-dimensional systems, for applications that need many reachable sets.

The first such case study is single-track vehicle model with seven states, presented in \cite{althoff2017commonroad}. The model takes into account tire slip making it a good candidate for studies that consider planning of evasive maneuvers that are very close to the physical limits. 
The system is described by the following hybrid $7$-dimensional non-linear single track (ST) model with parameters of a BMW $320$i car:
			
For $\vert x_4(k) \vert < 0.1$:
    \begin{align}\notag
    	\dot{x_i} &= a_i, \quad i \in \{1,\dots, 7\}\backslash\{3,4\}, \\\notag
    	\dot{x_3} &= \text{Sat}_1(p_1),\\\notag
    	\dot{x_4} &= \text{Sat}_2(p_2),
    \end{align}
and for $\vert x_4(k) \vert \ge 0.1$:
    \begin{align}\notag
    	\dot{x_i} &= b_i, \quad i \in \{1,\dots, 7\}\backslash\{3,4\}, \\\notag
    	\dot{x_3} &= \text{Sat}_1(p_1),\\\notag
    	\dot{x_4} &= \text{Sat}_2(p_2),
    \end{align}			

where,
\begin{align}\notag
	a_1 &=\! x_4\text{cos}(x_5),\quad a_2 = x_4\text{sin}(x_5),\quad a_5 = \frac{x_4}{l_{wb}} \text{tan}(x_3),\\\notag			
	a_6 &= \frac{p_2}{l_{wb}} \text{tan}(x_3) + \frac{x_4}{l_{wb}\text{cos}^2(x_3)} p_1, \quad a_7 = 0,\\\notag
	b_1 &=\! x_4\text{cos}(x_5+x_7),\quad b_2 =x_4\text{sin}(x_5+x_7),\quad b_5 = x_6,\\\notag
	b_6 &=\! \frac{\mu m}{I_z(l_r\!+\!l_f)}(l_fC_{S,f}(gl_r\!-\!p_2(k)h_{cg})x_3\!+\!(l_rC_{S,r}(gl_f\!+\!p_2h_{cg})\!-\!l_fC_{S,f}(gl_r\\\notag
	&~~~-p_2h_{cg}))x_7\!-\!(l_f^2C_{S,f}(gl_r\!-\!p_2h_{cg})\!+\!l_r^2C_{S,r}(gl_f\!+\!p_2h_{cg}))\frac{x_6}{x_4}),\\\notag
	b_7 &=\! \frac{\mu_f}{x_4(l_r\!+\!l_f)}(C_{S,f}(gl_r-p_2h_{cg})x_3\!+\!(C_{S,r}(gl_f+p_2h_{cg})\!+\!C_{S,f}(gl_r\\\notag
	&~~~-p_2h_{cg}))x_7\!-\!(l_fC_{S,f}(gl_r-p_2h_{cg})\!-\!l_rC_{S,r}(gl_f\!+\!p_2h_{cg}))\frac{x_6}{x_4})\\\notag
	&~~~-x_6.
\end{align}

\begin{wrapfigure}{r}{0.3\textwidth}
    \vspace{-1.0cm}
    \begin{center}
        \includegraphics[width=0.27\textwidth]{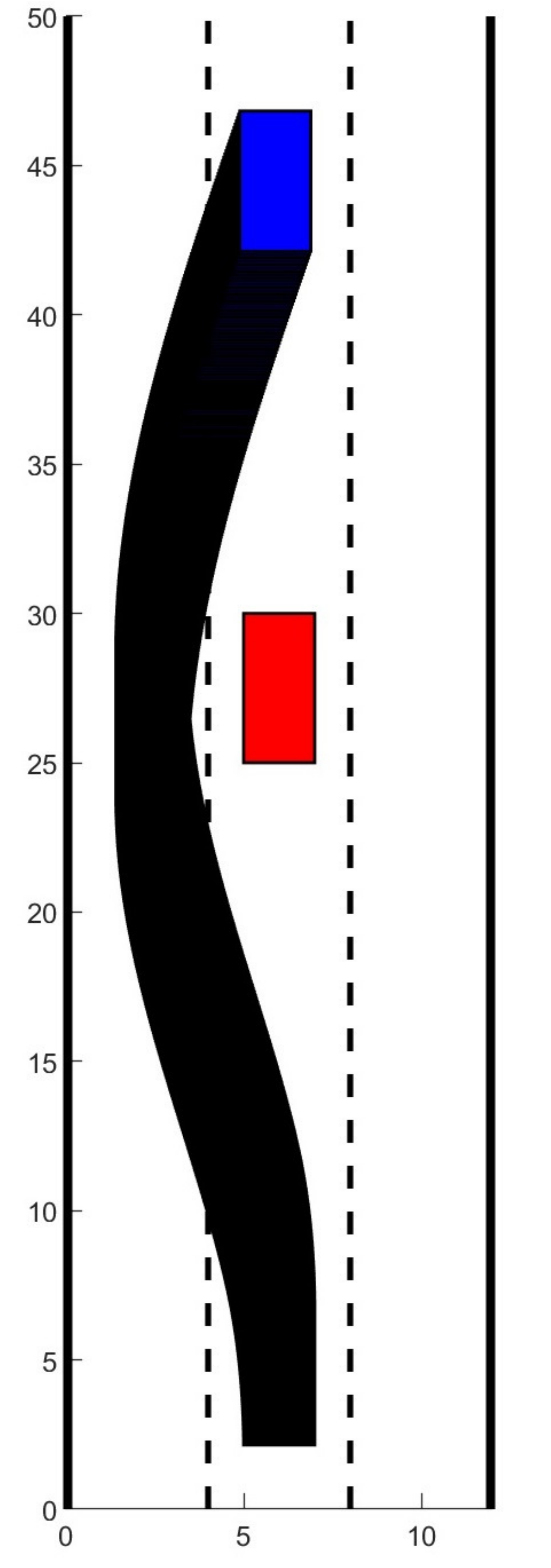}
        \caption{Reachable tube for the 320i car.}
        \label{fig:vehicle7d_lane_change}
    \end{center}
    \vspace{-1.0cm}
\end{wrapfigure}

Here, $\text{Sat}_1(\cdot)$ and $\text{Sat}_2(\cdot)$ are input saturation functions that are introduced in \cite[Section 5.1]{althoff2017commonroad},
$x_1$ and $x_2$ are the position coordinates, 
$x_3$ is the steering angle, 
$x_4$ is the heading velocity, 
$x_5$ is the yaw angle, 
$x_6$ is the yaw rate, and 
$x_7$ is the slip angle. 

Variables $p_1$ and $p_2$ are inputs and they control the steering angle and heading velocity, respectively. 
The following parameters for the BMW $320$i car: 
$l_{wb} = 2.5789$ as the wheelbase,
$m = 1093.3$ [kg] as the total mass of the vehicle,
$\mu = 1.0489$ as the friction coefficient, 
$l_f = 1.156$ [m] as the distance from the front axle to center of gravity (CoG), 
$l_r = 1.422$ [m] as the distance from the rear axle to CoG, 
$h_{cg} = 0.6137$ [m] as the hight of CoG, 
$I_z = 1791.6$ [kg m$^2$] as the moment of inertia for entire mass around $z$ axis, 
$C_{S,f} = 20.89$ [$1$/rad] as the front cornering stiffness coefficient, and 
$C_{S,r} = 20.89$ [$1$/rad] as the rear
cornering stiffness coefficient.

We consider a step-size of $0.005$ seconds between between the reach sets in a time window of $6.5$ seconds.
This results in computing $1300$ reach sets of the system.
We are interested in verifying the autonomous operation of the vehicle. 
We fix an input that performs a maneuver to overtake a car in the middle lane of a 3-lane highway.
We ran PIRK with the growth-bound method and extracted the reach pipe to verify the maneuver is applied successfully.
Within an Intel i7 processor with 8GB of RAM (Windows OS), PIRK managed to compute the reach pipe in 0.25 seconds. 
The reach pipe is plotted using the PIRK interface for MATLAB and it is shown in Fig. \ref{fig:vehicle7d_lane_change}.

\subsection{Performance on ARCH Benchmarks}

In order to compare \PIRK's performance to existing toolswt, we tested \PIRK's growth bound implementation on three systems from the ARCH-COMP'18 category report for systems with nonlinear dynamics \cite{immler2018arch}. This report contains benchmark data from several popular reachability analysis tools (\texttt{C2E2}, \texttt{CORA}, \texttt{Flow$^*$}, \texttt{Isabelle}, \texttt{SpaceEx}, and \texttt{SymReach}) on several nonlinear reachability problems with state dimensions between 2 and 12. 

\begin{table}
    \begin{tabular}{@{}rccccccc@{}}
    \toprule
    Benchmark model & \PIRK & \texttt{CORA} & \texttt{CORA/SX} & \texttt{C2E2} & \texttt{Flow$^*$} & \texttt{Isabelle} & \texttt{SymReach}
    \\\midrule
    Van der Pol (2 states) & 0.13 & 2.3 & 0.6 & 38.5 & 1.5 & 1.5 & 17.14 \\
    Laub-Loomis (7 states) & 0.04 & 0.82 & 0.85 & 0.12 & 4.5 & 10 & 1.93 \\
    Quadrotor (12 states) & 0.01 & 5.2 & 1.5 & - & 5.9 & 30 & 2.96 \\ \bottomrule
    \end{tabular}
    \label{tab:arch_times}
    \caption{Results from running \PIRK{} (growth bound method) to compute the reach sets for the examples reported in the ARCH-2018 competition.}
\end{table}

Table \ref{tab:arch_times} compares the computation times for \PIRK{} on the three systems are to those in the report. 
The results are compared to the times reported by other tools in \cite{immler2018arch}. 
All times are in seconds. 
\PIRK{} ran on an i9 CPU, while the others ran on i7 and i5: see \cite{immler2018arch} for more hardware details.

\PIRK{} solves each of the benchmark problems faster than the other tools. 
Both of the used i7 and i9 processors have multi processors (from 6 cores to 8 cores).
The advantage of \PIRK{} is its ability to utilize all available cores.

\section{Conclusion}
Using a simple parallelization of interval reachability analysis techniques, \PIRK{} is able to compute reachable sets for nonlinear systems faster and at higher dimensions than many existing tools. This performance increase comes from \PIRK{}'s ability to use massively parallel hardware such as GPUs and CPU clusters, as well as the use of parallelizable simulation-based methods. Of the three methods implemented, the growth bound method was the most efficient. Future work will focus on improving the memory-usage of the mixed monotonicity method and the Monte-Carlo based methods.

Future work will focus on two main directions: 
(1) improving the memory-usage of the mixed monotonicity method and the Monte-Carlo based method, and
(2) using \PIRK{} as a helper tool to synthesize controllers for high-dimensional systems.

\section*{Acknowledgements}

{\raggedright
This work was supported in part by the grants ONR N00014-18-1-2209, AFOSR
FA9550-18-1-0253, NSF ECCS-1906164.
}

\printbibliography[]

\end{document}